\documentclass[acmsmall]{acmart}

\AtBeginDocument{%
  }

\begin{document}

\setcopyright{cc}
\setcctype{by}
\acmJournal{PACMHCI}
\acmYear{2026}
\acmVolume{10}
\acmNumber{8}
\acmArticle{ISS030}
\acmMonth{11}
\acmDOI{10.1145/3859342}

\title{Sensory Comfort Across Multisensory XR Prototypes for Autistic Adults}

\author{Yibo Meng}
\orcid{0009-0009-0370-5456}
\authornote{Yibo Meng and Xiujun Chen contributed equally to this work.}

\affiliation{%
  \institution{Cornell University}
  \city{New York}
  \country{United States}}
\email{yibom2@illinois.edu}

\author{Xiujun Chen}
\authornotemark[1]
\orcid{0009-0006-5778-0345}
\affiliation{%
  \department{University College London}
  \institution{Bartlett School of Architecture}
  \city{London}
  \country{United Kingdom}}
\email{zcfthev@ucl.ac.uk}

\author{Bingyi Liu}
\orcid{0009-0000-4312-153X}
\affiliation{%
  \institution{University of Michigan}
  \city{Ann Arbor}
  \state{Michigan}
  \country{United States}}
\email{bingyi.ux@gmail.com}

\author{Xiaolan Ding}
\authornote{Corresponding author.}
\orcid{0009-0006-9956-4560}
\affiliation{%
  \institution{North China University of Science and Technology Health Science Center}
  \city{Tangshan}
  \country{China}}
\email{13838038961@163.com}

\renewcommand{\shortauthors}{Yibo Meng, Xiujun Chen, Bingyi Liu, and Xiaolan Ding}

\begin{abstract}
Multisensory XR may support autistic adults, yet sensory-friendly
design often prioritizes less input. We conducted a
mixed-methods, counterbalanced within-subjects study with 20 Chinese
autistic adults evaluating five complete XR configurations presenting
identical science content: text (L1), text with images (L2), click-triggered
spatial audio (L3), gradual light and audio (L4), and L4 plus vibrotactile
and thermal cues (L5). The measure combined comfort (Q1), absence of
sensory discomfort (Q2), and liking (Q3). Evaluations differed
(Friedman $\chi^2$(4) = 46.39, $p < .001$, Kendall's $W = .58$),
and a Q1--Q2 analysis retained the omnibus difference.
L3 was lowest ($M=2.97/5$; tied lowest for 16/20).
L5 was highest descriptively ($M=4.43/5$), but corrected comparisons
did not distinguish it from L1 or L4, and ceiling compression constrained
superiority claims.
Participants valued clarity, self-pacing, gradual transitions, exit
options, and coordinated cues, while disliking abruptness, simultaneous
demands, and insufficient warning.
These configuration-level findings identify participant-described
dimensions for controlled studies.
\end{abstract}




\maketitle

\section{Introduction}

Museums and science centers increasingly use XR to make scientific
phenomena concrete and situated. For autistic adults,\footnote{We use
identity-first language (``autistic adults'') because many English-speaking
autistic community members prefer it~\cite{kenny2016terms}, and because
guidance cautions against deficit-focused language~\cite{bottema2021language}.
The study was conducted in Chinese using plain-language terms that
participants confirmed they understood and accepted.} multisensory XR
requires particular attention: sensory hyper- and hypo-reactivity are part
of diagnostic criteria and are widely documented~\cite{apa2013dsm5,
robertson2017sensory,bensasson2009meta,crane2009sensory}.

XR can provide concrete, repeatable, and controllable experiences
~\cite{parsons2011vr,mesagresa2018vr,newbutt2020hmd}, but additional
channels may also create unwanted stimulation or divided attention
~\cite{robertson2017sensory,bensasson2009meta,tavassoli2018reactivity}.
Sensory-friendly initiatives often reduce ambient input
~\cite{manning2023sensoryinclusive}, whereas participatory work describes
enabling environments through interacting factors such as sensoryscape,
predictability, adjustment, and recovery~\cite{maclennan2023spiderweb}.
Thus, reduction is one accommodation strategy, not a sufficient rule for
evaluating complete XR designs that also differ in timing, interaction,
pacing, and opportunities to disengage.

Research on intolerance of uncertainty and predictive processing provides
interpretive vocabulary for timing, expectation, and control
~\cite{boulter2014uncertainty,neil2016uncertainty,
pellicano2012hypopriors,vandecruys2014hippea,sinha2014prediction}. These
contested accounts are background rather than tested mechanisms.

We conducted a counterbalanced within-subjects study with 20 Chinese
autistic adults who experienced five complete XR configurations (L1--L5)
presenting the same popular-science content. L1 presented text; L2
added images; L3 paired text and images with click-triggered spatial audio;
L4 used gaze or proximity to initiate gradual light and audio; and L5 added
vibrotactile and thermal cues to L4. These configurations jointly
varied in channels, trigger, onset, and exit affordances. Each was therefore
evaluated as a complete design, not as a controlled amount of sensory input
or an isolated test of any one modality. We asked:

\begin{itemize}
  \item \textbf{RQ1:} How does sensory comfort differ across five complete
  XR configurations for autistic adults?
  \item \textbf{RQ2:} How do autistic adults describe the qualities they
  valued or disliked in these configurations?
\end{itemize}

The five configurations received different evaluations: L3 was rated lower
than the others, while L4 and L5 received more positive evaluations in
several corrected comparisons. However, corrected comparisons
did not detect differences between L5 and L1 or L4, and L5 ratings were
compressed at the scale ceiling. Participants associated valued or disliked experiences
with clarity, self-pacing, abruptness, fragmentation, disengagement, gradual
transitions, and coordinated narrative cues.

This paper contributes (1) a counterbalanced within-subjects comparison of
five complete XR configurations, (2) a participant-grounded account of
valued and disliked sensory expressions, and (3) configuration-level design
considerations and candidate dimensions for future controlled studies.

\section{Related Work}

\subsection{Sensory Processing and Uncertainty in Autism}

Autistic sensory differences are recognized in diagnostic criteria and
span auditory, visual, tactile, and other modalities~\cite{apa2013dsm5,
bensasson2009meta,robertson2017sensory}. Hyper-reactivity,
hypo-reactivity, sensory seeking, and mixed profiles make any fixed
``accessible'' profile inadequate~\cite{crane2009sensory,
tavassoli2016sensory,tavassoli2018reactivity}.

Research on intolerance of uncertainty links sensory sensitivity,
predictability, and anxiety~\cite{boulter2014uncertainty,
neil2016uncertainty}; predictive-processing accounts similarly foreground
expectation and prediction formation~\cite{pellicano2012hypopriors,
vandecruys2014hippea,sinha2014prediction}. We use these contested accounts
as vocabulary for participant descriptions, not as explanations of our
results. They also expose a design gap: interface research has rarely
examined temporal predictability alongside modality and intensity.

\subsection{XR and Multisensory Technology for Autistic People}

XR research with autistic people emphasizes structured, repeatable, and
adjustable environments~\cite{parsons2011vr,mesagresa2018vr}, including
social training~\cite{kandalaft2013vr,didehbani2016vr} and school-based
head-mounted displays~\cite{newbutt2020hmd}. ISS systems likewise include
an eye-contact game and an interactive sensory VR environment
~\cite{xie2020maskme,keys2024anxiety}. Much of this work is
intervention-oriented and focuses on younger participants.

Critical HCI cautions against positioning autistic people primarily as
targets for normalization~\cite{spiel2019agency}; participatory research
instead emphasizes autistic expertise and involvement
~\cite{nicolaidis2019aasPIRE,fletcherwatson2019participation}. Our study
adopts a non-interventional orientation: it asks autistic adults what they
found comfortable or uncomfortable in informal learning prototypes.

ISS also offers precedents for learning across interactive furniture,
portable installations, mixed-reality cards, museum artifacts and displays,
and VR narratives~\cite{corona2016funf,gelsomini2018nomadic,
giraudeau2019cards,blumenstein2018interweaving,soni2020spherical,
lake2022space}. Together, this literature offers accommodation
principles and multisensory system examples, but limited within-participant
evidence about how autistic adults evaluate complete informal-learning XR
designs. We address that gap through participant-centered comparison
of five concrete configurations. Our contribution is
configuration-level evidence and a set of participant-described candidate
dimensions, not a general rule about the number of sensory channels.

\subsection{Multisensory Integration and Experiential Organization}

Multisensory HCI distinguishes meaningful cue combinations from decorative
additions~\cite{obrist2016touch}; crossmodal work describes temporal and
semantic relations across modalities~\cite{spence2011crossmodal}. Relevant
ISS systems span thermal displays, audio-tactile drawings,
auditory-vibrotactile museum experiences, haptic VR, and responsive
environments~\cite{kratz2017thermotouch,thevin2019accessible,
jung2019auditory,kumar2021touch,lyu2023social}. These studies establish a
broad design space, but not how autistic adults describe comfort in complete
multisensory configurations.

Feedforward supports anticipation~\cite{djajadiningrat2002donald,
wensveen2004frogger}, while sense-of-agency research relates control to the
fit among action and response~\cite{coyle2012agency,limerick2014agency}.
Gaze research identifies the ``Midas touch'' problem
~\cite{jacob1990gaze,majaranta2002eyetyping}, and ISS studies of learner
control and opt-in/out gestures emphasize how users enter and leave
interactions~\cite{karsten2017learner,rodriguez2017gesture}. We bring these
concerns together by examining participants' descriptions of timing, cue
relations, control, and individual sensitivity; each remains a candidate
dimension rather than a tested mechanism.

\section{Prototype Design}

\subsection{Content Materials}

All five prototype conditions presented the same set of seven short
popular-science content segments. Holding content constant across
conditions was the main control strategy: it prevented changes in topic,
wording, or narrative sequence from defining the five complete
configurations. The same seven-segment structure and narrative sequence
were used in each configuration.

One worked example was a sperm-whale diving segment. In the static
conditions, the segment appeared as text (L1) or text with image (L2).
In L3, selecting the segment enlarged the visual element and launched
the associated audio abruptly. In L4, approaching or dwelling on the
segment triggered a gradual visual-light transition and audio onset. In
L5, the same narrative event was layered with vibrotactile and thermal
cues: as the narration described descent into the deep sea, light
dimmed, the forearm temperature cue cooled, and the vibration followed
the described heartbeat rhythm. Across conditions, the underlying
science content and narrative order were held constant; only the
sensory channels, trigger, onset dynamics, and exit affordances varied
across the complete configurations.

\begin{figure*}[t]
  \centering
  \includegraphics[width=\textwidth]{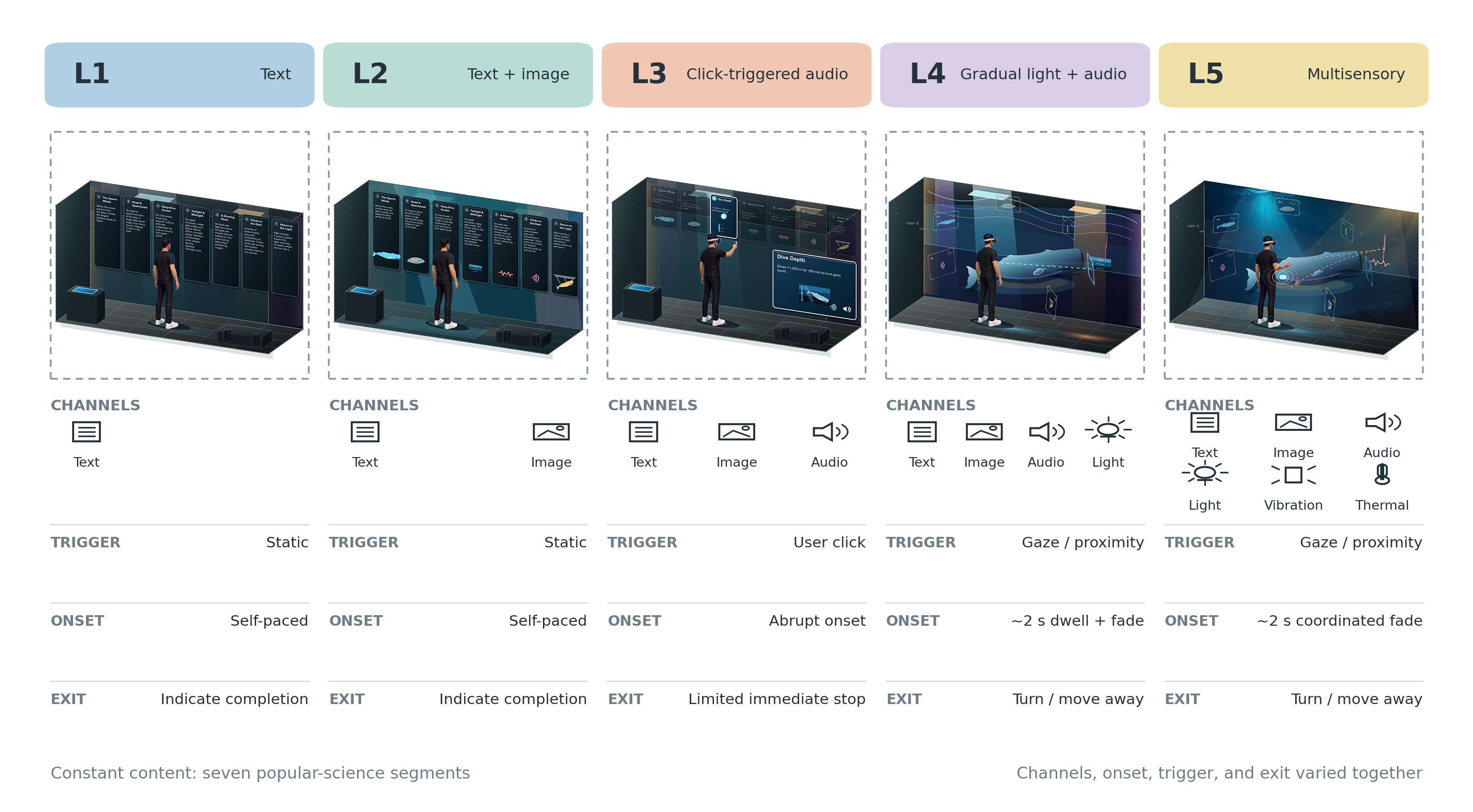}
  \caption{Overview of the five complete prototype
  configurations. Representative screenshots show what
  participants encountered; the lower rows summarize channels,
  trigger, onset, and exit structure. Content and narrative sequence
  remained constant while these design features varied together.}
  \Description{Five aligned columns labeled L1 through L5 show
  representative prototype screenshots and interaction summaries. L1
  displays text panels; L2 adds images; L3 shows a selected dive-depth
  segment with click-triggered audio; L4 shows a whale scene with gradual
  light and audio; and L5 shows the scene with additional vibration and
  thermal cues. Rows below each screenshot list channels, trigger
  type, onset, and exit mechanism.}
  \label{fig:prototype-overview}
\end{figure*}

Figure~\ref{fig:prototype-overview} presents representative screenshots
and summarizes the interaction structure of the five complete prototypes.

\subsection{Hardware and Sensory Setup}

The prototype used an XR headset for visual display and spatial audio,
and L5 additionally used a forearm-worn vibrotactile and thermal patch.
Before formal exposure, participants completed a neutral-scene adaptation
stage and could pause or withdraw at any point. A rest period of at least
five minutes separated successive conditions. Appendix~\ref{app:hardware}
reports the hardware and stimulus settings used to implement these cues.

\subsection{Prototype Configurations and Interaction Characteristics}

\begin{table*}
  \caption{Sensory and interaction characteristics of the five complete prototype configurations}
  \label{tab:condition-matrix}
  \begin{tabular}{p{.08\linewidth}p{.27\linewidth}p{.18\linewidth}p{.24\linewidth}p{.16\linewidth}}
    \toprule
    Condition & Channels & Trigger type & Onset envelope & User exit mechanism \\
    \midrule
    L1 & Text only & Static presentation & No automated sensory onset & Participant indicated completion \\
    L2 & Text + image & Static presentation & No automated sensory onset & Participant indicated completion \\
    L3 & Text + image + spatial audio & User click & Abrupt onset after click; no programmed fade-in & Segment-level audio could not be immediately interrupted once triggered \\
    L4 & Text + image + spatial audio + light & Gaze/proximity dwell & Approximately two-second dwell before gradual onset; gradual fade when gaze/proximity ended & Turning away or moving away initiated fade-out \\
    L5 & L4 channels + vibrotactile/thermal cues & Gaze/proximity dwell & Approximately two-second dwell before coordinated gradual onset across channels & Turning away or moving away initiated fade-out \\
    \bottomrule
  \end{tabular}
\end{table*}

The approximately two-second dwell threshold in L4 and L5 was chosen
as a pragmatic safeguard against accidental activation, reflecting the
well-known gaze-interaction problem that looking at an object does not
always mean intending to activate it~\cite{jacob1990gaze,
majaranta2002eyetyping}. L3 deliberately preserved a familiar
click-to-trigger interaction. Participants later mentioned its abrupt
onset and limited interruptibility in accounts of fragmentation and loss
of control.

\subsection{Rationale for the Complete Prototype Set}

The five conditions were developed by successively layering features to
create ecologically plausible exhibition prototypes, beginning with static
text and extending to multisensory presentation. This construction process
was not treated as a controlled manipulation of sensory quantity. The study
compared evaluations of the five prototypes while keeping the science
content stable and the session manageable for participants. Each L-level
served as a complete unit of experience: sensory channels, interaction
structure, onset, and exit together defined that level. The design supports
configuration-level comparison and participant accounts of valued or
disliked experiences, not claims about the independent effect of any one
component.

\section{Method}

\subsection{Participants}

\subsubsection{Recruitment Process}

This study recruited participants through online social media. The
research team posted recruitment notices in three autistic adult related
social media communities and under relevant hashtags. Inclusion
criteria were as follows. Participants had to be at least 18 years
old. They needed to hold an autism diagnosis certificate issued by a
formal medical or psychological assessment institution. They had to
be able to complete informed consent independently. They could not
have uncorrected severe visual or auditory impairment, since the
prototype experience involved visual imagery and spatial audio
presentation. They could not have a history of photosensitive
epilepsy or other contraindications for XR device use. Exclusion
criteria included the following. Participants with a clear prior
history of adverse reactions to head mounted displays, vibration
patches, or temperature patches were excluded. Those currently in an
acute mental health crisis were excluded. Those unable to understand
experimental instructions or independently express their sensory
experience were also excluded.

\subsubsection{Recruitment Results}

The recruitment received a total of 25 sign up submissions. After
initial screening, which included verification of diagnosis
certificates, confirmation of age range, and checking for device use
contraindications, 20 participants ultimately met all inclusion
criteria and completed informed consent, entering the formal
experimental process. Participant ages ranged from 19 to 55 years,
with 10 males and 10 females, an even distribution. Ten participants
held a bachelor's degree, six held a high school education, three
held a middle school education, and one held only an elementary
school education, covering a fairly wide range from elementary
school to bachelor's degree. The sample showed a degree of
heterogeneity in age, gender, and education level. Detailed
information is shown in Table~\ref{tab:participants}.

\begin{table}
  \caption{Participant demographics}
  \label{tab:participants}
  \begin{tabular}{cccl}
    \toprule
    ID & Age & Gender & Education \\
    \midrule
    P1  & 22 & M & Bachelor's \\
    P2  & 34 & F & Bachelor's \\
    P3  & 36 & M & High school \\
    P4  & 45 & F & High school \\
    P5  & 39 & M & High school \\
    P6  & 55 & F & Middle school \\
    P7  & 19 & M & High school \\
    P8  & 21 & M & Bachelor's \\
    P9  & 24 & F & Bachelor's \\
    P10 & 25 & F & Bachelor's \\
    P11 & 36 & M & Middle school \\
    P12 & 37 & F & Middle school \\
    P13 & 44 & M & High school \\
    P14 & 34 & M & Bachelor's \\
    P15 & 32 & M & Bachelor's \\
    P16 & 20 & F & High school \\
    P17 & 37 & F & Bachelor's \\
    P18 & 41 & F & Bachelor's \\
    P19 & 38 & F & Elementary school \\
    P20 & 29 & M & Bachelor's \\
    \bottomrule
  \end{tabular}
\end{table}

\subsubsection{Ethics}

This study protocol was reviewed and approved by the ethics committee at
the authors' organization, and the entire process
followed the Declaration of Helsinki and relevant national human
research ethics standards, giving full consideration to the
communication and accessibility needs of autistic participants regarding informed consent,
communication, and sensory safety. Before formal participation, all
participants received a one on one explanation of informed consent,
using plain language supplemented with illustrations to explain the
research purpose, procedure, and the types of sensory stimuli that
might be involved, ensuring that participants could decide whether
to participate based on full understanding. Participants were
clearly informed that they could pause or withdraw from the
experiment at any time without giving a reason. Before device use,
there was a neutral scene adaptation stage, used to exclude cases
showing clear resistance to the headset or the vibration and
temperature patches. A mandatory rest period followed each
condition, and participants could choose to extend their rest time
on their own. All interview recordings, scale data, and diagnostic
information were anonymized, with participant ID numbers replacing
real names, and the raw materials were used only within the research
team. Each participant received compensation of 40 RMB
(approximately 6 US dollars). Given the total participation time of
approximately 90 to 110 minutes, this compensation corresponded to
roughly 22 to 27 RMB per hour and was approved as part of the ethics
review.

\subsection{Intervention Procedure}

This study completed all data collection through a single laboratory
visit, with each participant's total participation time controlled
to approximately 90 to 110 minutes, in order to avoid sensory
fatigue or attention decline in autistic participants during long
tasks. The overall procedure was divided into four stages. (1) Pre
test preparation and adaptation. (2) Sequential experience of the
five interaction conditions with immediate measurement. (3) A
retrospective interview after all conditions were completed. (4)
Conclusion and exit interview. The specific procedure is described
below.

\begin{figure*}[t]
  \centering
  \includegraphics[width=\textwidth]{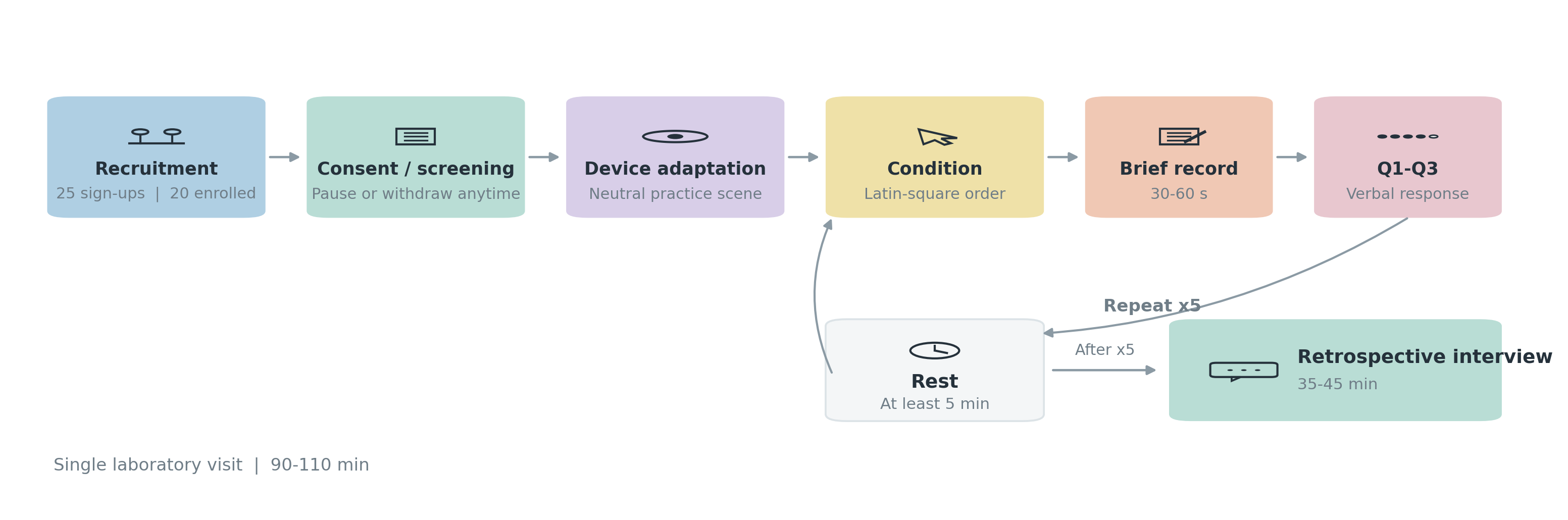}
  \caption{Study procedure. Participants completed consent, screening,
  and device adaptation before five Latin-square-ordered conditions.
  Each condition was followed by an immediate record, Q1--Q3 ratings,
  and at least five minutes of rest; a retrospective interview followed
  the five-condition loop.}
  \Description{A horizontal flow diagram from recruitment through
  consent and screening, device adaptation, condition exposure, a brief
  record, and Q1 through Q3 ratings. Ratings lead to a rest period that
  loops back to the condition stage for five conditions. The flow then
  ends with a 35 to 45 minute retrospective interview.}
  \label{fig:procedure}
\end{figure*}

Figure~\ref{fig:procedure} shows the single-visit study flow and the
condition-level measurement loop.

\subsubsection{Preparation}

After participants arrived at the laboratory, the researcher first
introduced the overall arrangement of the experiment using plain and
specific language, including the number of prototypes to be
experienced, the approximate duration of each prototype, and the
right to pause or withdraw at any time. The researcher specifically
noted that there were no right or wrong responses in this
experiment, and that the purpose was simply to understand the
participant's genuine feelings about different presentation methods,
in order to reduce evaluation anxiety.

This was followed by a device adaptation stage. Participants wore
the XR headset (and, for the L5 condition, the forearm vibration and
temperature patch device), and before the formal experience began,
they entered a neutral practice scene unrelated to the seven
segments of popular science content, practicing basic gaze,
pointing, and click operations, until they confirmed that they could
perform the interaction actions comfortably and showed no discomfort
from wearing the device itself.

\subsubsection{Experiment}

Each participant experienced the five prototypes L1 through L5 in
sequence, following the 5 by 5 Latin square order corresponding to
their assigned group (Group A through E). The experience procedure
for each condition was as follows.

\begin{enumerate}
  \item \textbf{Condition presentation.} The researcher briefly
  informed the participant of the number of the prototype about to
  begin, without revealing the specific design differences, in order
  to avoid expectation bias, and the system then launched the
  corresponding condition. L1 and L2 were static presentations, in
  which participants could freely control their reading pace with no
  fixed minimum duration, and the system prompted ``you can tell me
  once you feel you have finished viewing.'' L3 was a click
  triggered condition, in which participants could freely decide
  which passages to click, the order of clicking, and how many times
  to repeat a click, with no requirement to trigger all seven
  segments. L4 and L5 were gaze or proximity triggered conditions, in
  which participants could move or turn freely within the space, and
  the trigger rule was that content would play automatically after
  about two seconds of sustained gaze or proximity to a given
  segment.

  \item \textbf{Immediate brief record.} After each condition ended,
  the researcher conducted a brief interview lasting about 30 seconds
  to 1 minute, serving as an anchor style record for that condition,
  structured mainly around two open questions: ``Was there any moment
  during this experience that made you feel uncomfortable? Can you
  tell me specifically when that was?'' The purpose of this stage was
  not to obtain a complete narrative, but to preserve a raw
  impression anchor while the participant's memory was still fresh,
  to be used later to prompt recall during the retrospective
  interview. The duration was therefore strictly controlled, to
  avoid excessively interrupting the continuity of the experience or
  adding to the participant's overall task load.

  \item \textbf{Completion of the sensory comfort scale.} After the
  immediate brief record, participants completed the sensory comfort
  scale for that condition (Q1: after experiencing this system, my
  overall feeling is comfortable; Q2: the sensory stimuli in this
  system did not make me feel uneasy or uncomfortable; Q3: overall, I
  liked this experience), with the researcher reading each item aloud
  and recording the participant's verbal response.

  \item \textbf{Rest between conditions.} A mandatory rest period of
  at least 5 minutes was set after each condition, during which
  participants could choose to remove the device, leave the
  experience area, or rest quietly.
\end{enumerate}

\subsubsection{Retrospective Interview}

After all five conditions had been experienced and the immediate
records and scale responses had been completed, participants entered
the retrospective interview stage, which lasted approximately 35 to
45 minutes. In this stage, the researcher presented each participant
with their own immediate brief records from the five conditions, one
at a time, as memory prompts, guiding them to review each one in the
order they had actually experienced the conditions. The researcher
then focused on comparative questions across conditions, for
example: ``You mentioned earlier that both L3 and L1 had quiet
beginnings, but they felt different. Can you tell me specifically
what felt different?'' ``Among these five experiences, which one
made you feel most relaxed? Why?'' ``Was there any experience that
felt less tiring to you even though it had richer content?'' The
follow up questioning followed a principle of open ended questions
first, followed by gradual focusing. The researcher first allowed
participants to describe their experience on their own, avoiding
leading questions with a predetermined conclusion, and only pursued
specific details once participants had spontaneously mentioned a
relevant experience.

\subsection{Data Analysis Method}

This study adopted an analysis strategy combining quantitative and
qualitative methods. The former tested differences in sensory comfort
among the five complete interaction prototypes, while the latter identified
qualities that participants associated with valued or disliked experiences
in those prototypes. Integration maintained the distinction among direct
quantitative findings, participant descriptions, and candidate questions
for future controlled work; themes were not treated as measured mediators.
The integration logic followed mixed-methods guidance that treats
joint interpretation, rather than parallel reporting alone, as the key
analytic move~\cite{fetters2013integration,creswell2018mixed}.

\subsubsection{Quantitative Data Analysis Method}

The quantitative analysis used the 300 item-level scores in the source
workbook (20 participants by five conditions by three items). There
were no missing ratings. We calculated Cronbach's alpha for Q1
(overall comfort), Q2 (absence of sensory discomfort), and Q3
(overall experience preference) within each condition and after
stacking the five condition records. The pooled 95\% confidence
interval was estimated with a 20,000-iteration percentile bootstrap
that resampled participants as clusters, preserving the five repeated
condition records within each resample. Each participant's Q1--Q3
scores were then averaged to produce one composite score per condition.
We retain the study protocol's term \emph{sensory comfort scale} for
this measure. However, because Q3 asks about overall liking rather
than comfort directly, the Q1--Q3 composite is interpreted as a mixed
condition-level evaluation, not as a pure measure of sensory comfort.
The Q1--Q2 analysis provides the narrower comfort-focused sensitivity
check.

Because the dependent measure was a three-item Likert composite and L5
departed from normality, the Friedman test was the primary omnibus
analysis~\cite{norman2010likert,sullivan2013likert}. Kendall's $W$
quantified the omnibus effect. A repeated-measures ANOVA is retained
only as a scale-level sensitivity analysis, with both partial and
generalized eta squared reported. Shapiro--Wilk tests are provided as
distributional descriptions rather than as a gatekeeping rule
~\cite{shapiro1965normality}.

After the omnibus test, all ten two-sided Wilcoxon signed-rank
comparisons were calculated~\cite{wilcoxon1945ranking}. The normal
approximation included tie handling and a continuity correction; zero
differences were omitted from the rank calculation. We report $W$,
$z$, $r=|z|/\sqrt{n_{\mathrm{nonzero}}}$, and the paired
Hodges--Lehmann location estimate. Bonferroni adjustment was applied to
the family of ten comparisons (equivalently, corrected
$\alpha=.005$).

Three robustness analyses examined whether the omnibus result depended
on measurement or individual cases: separate Friedman tests for Q1,
Q2, and Q3; a two-item composite restricted to Q1 and Q2; and
leave-one-participant-out Friedman tests.

Two additional exploratory diagnostics examined the internal structure
of the condition pattern. First, we decomposed the adjacent L1--L2 and
L2--L3 contrasts by item, giving six Wilcoxon tests in one Bonferroni
family. These tests were post hoc and are used to localize score changes,
not to validate three independent subscales. Second, we counted
participant-level directions and ties for four descriptive patterns:
whether L3 was a participant's lowest condition, whether L5 was highest,
whether L4 exceeded L3, and whether L5 exceeded L1. Direction counts are
reported without population-level inference.

As a secondary exploratory diagnostic, we grouped the prototypes into
the static configurations (mean of L1 and L2), L3, and the coordinated
multisensory configurations (mean of L4 and L5), then applied a Friedman
test and three Bonferroni-adjusted Wilcoxon comparisons. Because each
family pools multidimensional configurations, the five complete L-levels
remain the primary units of analysis.

\subsubsection{Qualitative Data Analysis Method}

The qualitative data analysis in this study used reflexive thematic
analysis~\cite{braun2006thematic,braun2019reflecting} to code the complete
transcripts formed by the 20 participants across the immediate brief
interviews and the retrospective interviews, with the goal of
understanding which qualities participants valued or disliked in the five
complete configurations and how they described comfortable and
uncomfortable moments. The analysis did not treat interview themes as
causal explanations of the quantitative differences.

The analysis process followed six steps. The first step was
familiarization with the data. Two researchers each read through the
complete interview transcripts (including each participant's
immediate brief records under the five conditions and the
retrospective interview after all conditions were completed), while
repeatedly listening back to the original recordings, marking
initial impressions and noteworthy statements, and forming their own
initial notes. The second step was generating initial codes. The two
researchers independently performed open coding on the entire text,
using semantically complete segments of speech as the coding unit
rather than predetermined fixed categories, aiming to stay close to
the language and emotional tone of participants' original
statements. For example, the statement about the sound suddenly
appearing the instant a click occurred, and the feeling of a jolt in
the chest, was initially coded as ``suddenness of trigger'' and
``lack of warning signal.'' The third step was searching for themes.
The two researchers cross checked their respective coding
frameworks, merging and clustering codes with similar meanings to
form the outline of several candidate themes. At this stage, the
researchers noticed that three groups of codes initially labeled
separately as ``discomfort associated with suddenness,'' ``gradual change
described as buffering,'' and ``sense of control and expectation
management'' often appeared intertwined and mutually reinforcing in
participants' actual statements, suggesting that the three might
jointly point to the same deeper, more holistic experience. The
fourth step was reviewing themes. The researchers returned to the
original transcripts, checking each candidate theme against the data
itself, and examining whether the boundaries between themes were
clear and whether substantial overlap existed. After repeated
discussion, the three groups of codes above were merged into one
broader experiential theme, the fragmented feeling of L3, while
three other independent themes with good internal consistency were
retained (sensory simplicity, sensory unity, and the individualized
spectrum of sensory sensitivity). The fifth step was defining and
naming themes. Definitions were written for each of the four
retained themes, clarifying their core argument, boundaries, and
relationship to the other themes, and the most representative
quotations that accurately conveyed the meaning of each theme were
selected from the original transcripts as examples. The sixth step
was writing the analytic report, integrating the four themes and
their supporting quotations into a narrative results report and
cross referencing them with the quantitative results (such as the
main effect test of condition and participant-level direction counts),
placing participant descriptions alongside the quantitative configuration
comparisons without treating the themes as mediators or causal
explanations.

Throughout the analysis process, the two researchers consistently
resolved disagreements through negotiated consensus rather than by
calculating an inter coder reliability coefficient. This approach is
consistent with the methodological stance of reflexive thematic
analysis, in which themes are regarded as interpretive products
actively constructed by the researcher based on the data, rather
than fixed entities independent of the researcher that can be
objectively measured and tested for reliability. The researchers'
analytic background, including their familiarity with the literature
on autism sensory sensitivity and their prior knowledge of the
quantitative findings, was treated as an inseparable part of the
analysis process that should be openly acknowledged, rather than a
source of bias to be eliminated through procedural control.

Appendix~\ref{app:codebook} documents the final thematic framework,
including each theme's definition, indicative initial codes, analytic
boundaries, and associated configurations. The framework records the
outcome of the reflexive process rather than a codebook fixed before
analysis.

At the integration stage, participant identifiers were used to align
quoted accounts with their own five-condition score profiles. P3 was
examined because this participant's configuration profile differed from
the group-level ordering and also supplied a key L3 onset account; P11 and
P17 were examined because their interview accounts diverged from what a
single condition-level score or group pattern might imply. This case-level
integration was exploratory and is not treated as evidence that any one
profile represents a subgroup.

\subsubsection{Translation, Reporting, and Positionality}

All interviews were conducted in Chinese. Quotations used in this
paper were translated into English by a bilingual researcher and then
checked by a second bilingual researcher against the original Chinese
transcripts for conceptual equivalence rather than word-for-word
literalness. Reporting of qualitative procedures was guided by
established interview and qualitative reporting recommendations
~\cite{tong2007coreq,obrien2014srqr}.

The research team had HCI and XR design expertise and prior familiarity
with autism sensory literature, but the current prototype was not
co-designed with an autistic advisory group. This position gave the
team technical sensitivity to configuration design, temporal
organization, and sensory coherence, while also creating a risk that
participant accounts would be interpreted through design concepts too
quickly. During analysis, the team used memoing and repeated return to
participant language to keep sensory descriptions grounded in the
interview material. The absence of autistic co-researchers is treated
as a limitation and motivates future participatory work.

\section{Results}

The quantitative results section (Section~\ref{sec:quant}) reports
the differences in sensory comfort among the five interaction
prototypes and their statistical tests. The qualitative results
section (Section~\ref{sec:qual}) presents the experiential themes
constructed through interview analysis. Finally,
Section~\ref{sec:integration} aligns the direct quantitative findings with
participant descriptions and states bounded candidate questions.

\subsection{Quantitative Results}
\label{sec:quant}

\subsubsection{Reliability Test Results}

All 300 item responses were present. After stacking the five repeated
condition records, Cronbach's alpha was .895; the participant-cluster
bootstrap 95\% CI was [.850, .925]. Condition-specific point estimates
ranged from .782 to .905 (Table~\ref{tab:alpha}). Their bootstrap
intervals were wide, as expected with 20 participants, so the point
estimates support item coherence but do not establish construct purity
within every condition.

\begin{table}
  \caption{Cronbach's alpha for Q1--Q3 with participant-cluster bootstrap intervals}
  \label{tab:alpha}
  \begin{tabular}{lcc}
    \toprule
    Scores & $\alpha$ & 95\% CI \\
    \midrule
    L1 & .905 & [.706, .965] \\
    L2 & .855 & [.700, .923] \\
    L3 & .782 & [.528, .892] \\
    L4 & .802 & [.574, .904] \\
    L5 & .850 & [.522, .930] \\
    All condition records & .895 & [.850, .925] \\
    \bottomrule
  \end{tabular}
\end{table}

\subsubsection{Descriptive Statistics}

\begin{figure*}[t]
  \centering
  \includegraphics[width=\textwidth]{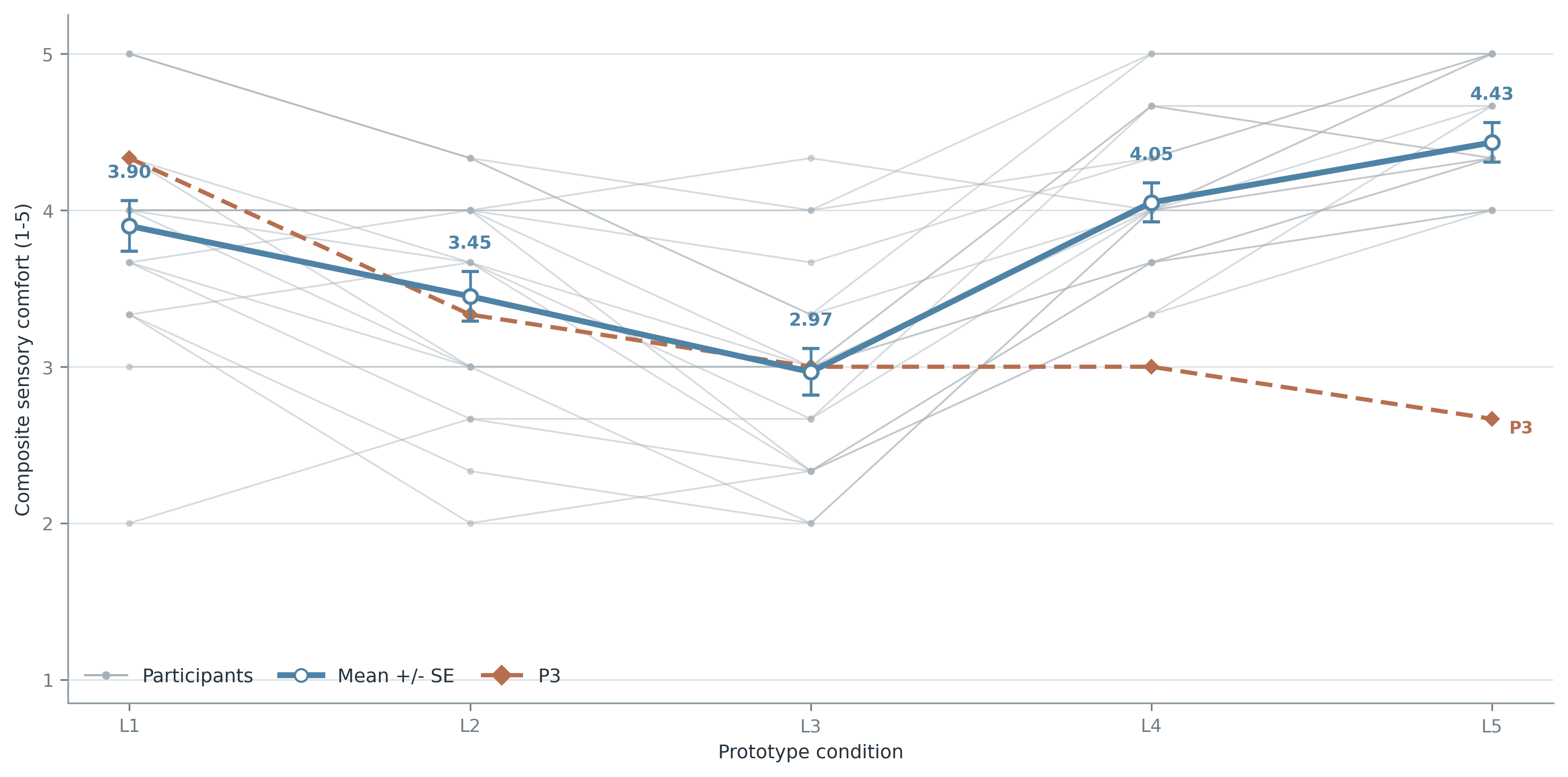}
 
  \caption{Participant-level Q1--Q3 composite sensory comfort profiles across configurations. Gray lines show individual composite scores, the blue line shows the mean with $\pm$1 standard error, and P3's profile is highlighted.}
  \Description{A paired profile plot for 20 participants across L1 through L5. Most participants give L3 a lower score than L4 and L5; the group mean is lowest at L3 and highest at L5. P3 is highlighted because its configuration profile differs from the group-level ordering.}
  \label{fig:means}
\end{figure*}

Descriptive statistics for the protocol-defined composite sensory
comfort scores are shown in Table~\ref{tab:descriptives}. This Q1--Q3
composite combines two comfort-focused items with one overall-liking item
and is therefore interpreted as a mixed condition-level evaluation. Across the
five configurations, L3 had the lowest mean and L5 the highest
(Figure~\ref{fig:means}). These are descriptive positions;
inferential comparisons and the narrower Q1--Q2 sensitivity analysis are
reported below.

\begin{table}
  \caption{Descriptive statistics of sensory comfort across the five conditions}
  \label{tab:descriptives}
  \begin{tabular}{lcccc}
    \toprule
    Condition & M & SD & Median & $Q_1$--$Q_3$ \\
    \midrule
    L1 (text only) & 3.90 & 0.72 & 4.00 & 3.58--4.33 \\
    L2 (+ image) & 3.45 & 0.70 & 3.67 & 3.00--4.00 \\
    L3 (+ sound, click triggered) & 2.97 & 0.67 & 3.00 & 2.33--3.33 \\
    L4 (+ light, gradual triggered) & 4.05 & 0.55 & 4.00 & 3.67--4.42 \\
    L5 (+ vibration/temperature, gradual triggered) & 4.43 & 0.56 & 4.33 & 4.25--5.00 \\
    \bottomrule
  \end{tabular}
\end{table}

\subsubsection{Main Effect Test of Condition}

Shapiro--Wilk tests did not detect departures from normality for L1--L4
(all $p>.10$), whereas L5 departed from normality ($W=.818$,
$p=.002$). The primary Friedman test showed a condition difference,
$\chi^2(4)=46.39$, $p<.001$, Kendall's $W=.580$. The supporting
repeated-measures ANOVA led to the same omnibus conclusion,
$F(4,76)=31.43$, $p<.001$, partial $\eta_p^2=.623$ and generalized
$\eta_G^2=.394$. The latter two quantities are reported with explicit
labels because .394 is a generalized, not partial, effect size.

\subsubsection{Post Hoc Pairwise Comparisons}

The Bonferroni-adjusted Wilcoxon tests are shown in
Table~\ref{tab:pairwise}. L3 was lower than each of the other four
conditions. L2 was lower than L1, L4, and L5. We did not detect
corrected differences for L1--L4, L1--L5, or L4--L5. In particular,
the L1--L5 result ($p_{\mathrm{adj}}=.096$) is a failure to detect a
difference under this correction, not evidence that the two conditions
are equivalent.

\begin{table*}
  \caption{Wilcoxon signed-rank comparisons of composite comfort scores}
  \label{tab:pairwise}
  \begin{tabular}{lrrrrrl}
    \toprule
    Comparison & $W$ & $z$ & $p_{\mathrm{adj}}$ & $r$ & HL & Direction \\
    \midrule
    L1--L2 & 8.5  & -2.904 & .037 & .750 & 0.50 & L1 $>$ L2 \\
    L1--L3 & 4.5  & -3.527 & .004 & .831 & 1.00 & L1 $>$ L3 \\
    L1--L4 & 41.0 & -1.376 & 1.000 & .344 & -0.17 & n.s. \\
    L1--L5 & 14.0 & -2.591 & .096 & .669 & -0.50 & n.s. \\
    L2--L3 & 7.5  & -2.965 & .030 & .766 & 0.50 & L2 $>$ L3 \\
    L2--L4 & 11.0 & -3.082 & .021 & .748 & -0.67 & L4 $>$ L2 \\
    L2--L5 & 6.5  & -3.669 & .002 & .820 & -1.00 & L5 $>$ L2 \\
    L3--L4 & 1.5  & -3.752 & .002 & .861 & -1.17 & L4 $>$ L3 \\
    L3--L5 & 1.0  & -3.875 & .001 & .867 & -1.50 & L5 $>$ L3 \\
    L4--L5 & 17.0 & -2.618 & .088 & .654 & -0.33 & n.s. \\
    \bottomrule
  \end{tabular}
  \begin{minipage}{\linewidth}\footnotesize
  Note: $p_{\mathrm{adj}}$ is Bonferroni-adjusted across ten tests. HL is
  the paired Hodges--Lehmann estimate for the left condition minus the
  right condition. Effect size $r$ uses the number of nonzero pairs.
  \end{minipage}
\end{table*}

\subsubsection{Robustness, Item Decomposition, and Within-Participant Consistency}

The condition effect was present for each item analyzed separately:
Q1, $\chi^2(4)=56.36$, $p<.001$, $W=.704$; Q2,
$\chi^2(4)=33.61$, $p<.001$, $W=.420$; and Q3,
$\chi^2(4)=46.72$, $p<.001$, $W=.584$. Restricting the composite to
the two items that refer directly to comfort and discomfort (Q1 and
Q2) also retained the omnibus pattern, $\chi^2(4)=45.34$, $p<.001$,
$W=.567$. Thus, the condition effect was not produced solely by Q3's
preference wording, although individual pairwise conclusions were less
stable under the two-item scoring. Finally, omitting each participant
in turn produced Kendall's $W$ values from .560 to .684 (all
$p<.001$), indicating that the omnibus result was not dependent on one
participant.

Table~\ref{tab:item-means} decomposes the composite by item. The
exploratory L1--L2 composite difference of 0.45 was concentrated in Q1
(mean difference = 0.80, $p_{\mathrm{adj}}=.003$) and, descriptively,
Q3 (0.45, $p_{\mathrm{adj}}=.088$), rather than Q2 (0.10,
$p_{\mathrm{adj}}=1.000$). By contrast, the 0.48 decrease from L2 to L3
was largest for Q2 (0.85, $p_{\mathrm{adj}}=.030$); the Q1 and Q3
differences were 0.35 ($p_{\mathrm{adj}}=.251$) and 0.25
($p_{\mathrm{adj}}=.719$), respectively. Thus, the static L1--L2
difference was not carried by the item asking about sensory discomfort,
whereas the L2--L3 decrease was concentrated in that item. Because these
were post-hoc single-item tests, this localization should not be read as
evidence for distinct validated subscales.

\begin{table}
  \caption{Item-level mean scores across conditions}
  \label{tab:item-means}
  \begin{tabular}{lccc}
    \toprule
    Condition & Q1 comfort & Q2 no discomfort & Q3 liking \\
    \midrule
    L1 & 4.00 & 3.80 & 3.90 \\
    L2 & 3.20 & 3.70 & 3.45 \\
    L3 & 2.85 & 2.85 & 3.20 \\
    L4 & 3.95 & 4.05 & 4.15 \\
    L5 & 4.25 & 4.45 & 4.60 \\
    \bottomrule
  \end{tabular}
\end{table}

\begin{figure*}[t]
  \centering
  \includegraphics[width=\textwidth]{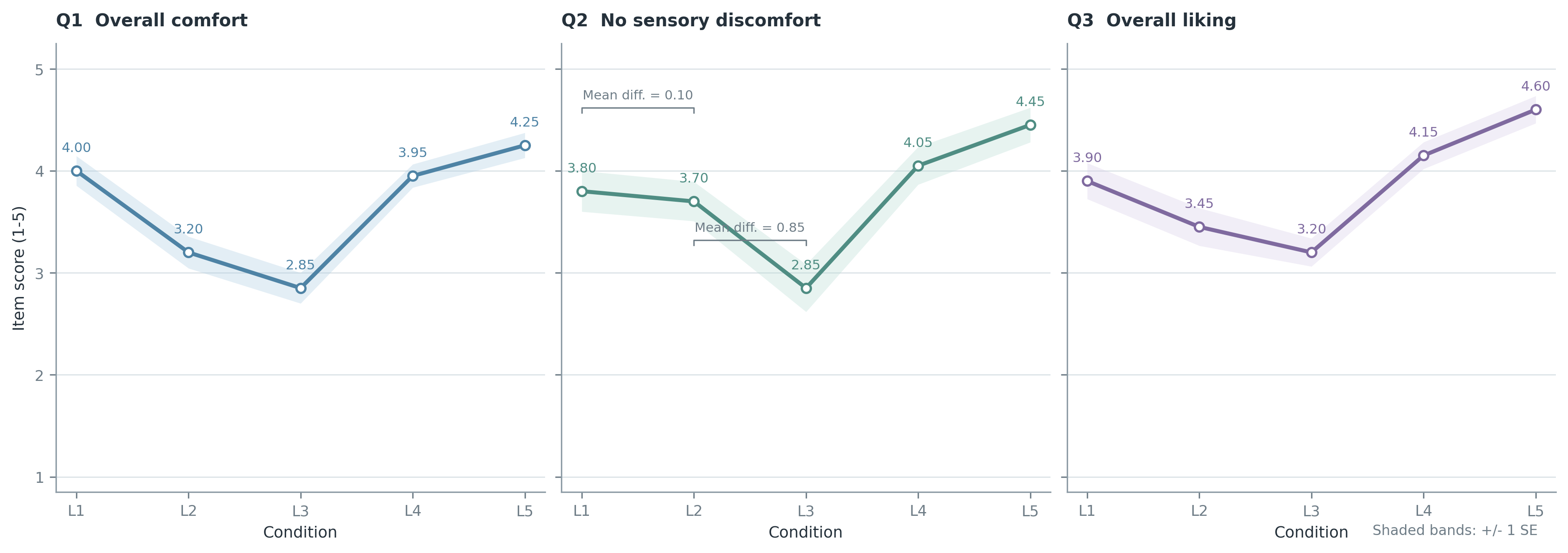}
  \caption{Item-level response patterns across the five conditions.
  Lines show item means and shaded bands show $\pm$1 standard error. The
  brackets in Q2 mark the descriptive L1--L2 and L2--L3 mean
  differences.}
  \Description{Three aligned line plots show means for overall comfort,
  absence of sensory discomfort, and overall liking. L3 has the lowest
  mean in all three plots, while L4 and L5 have higher means. In the no
  sensory discomfort panel, L1 and L2 differ by 0.10 while L2 and L3
  differ by 0.85.}
  \label{fig:item-patterns}
\end{figure*}

Figure~\ref{fig:item-patterns} visualizes the same item-level means with
standard-error bands, making the localization of the L2--L3 decrease in
Q2 visible alongside the configuration-level item differences.

The group ordering was also visible within participants. L3 was tied
for the lowest-scoring condition for 16 of 20 participants and was the
unique lowest for 11. L5 was tied for highest for 17 participants and
was the unique highest for 12. The L4 score exceeded L3 for 18
participants, with one tie and one reversal. These counts show that the
L3--L4 contrast was not produced by a small set of extreme profiles,
while retaining evidence of individual exceptions.

For L1--L5, 14 participants rated L5 higher, one rated L1 higher, and
five tied. This 14:1 directional imbalance does not override the
Bonferroni-adjusted inferential result ($p_{\mathrm{adj}}=.096$), but it
also makes clear that non-significance is not evidence of equal
responses. L5 scores were compressed at the upper boundary: 30 of 60
item responses (50\%) were 5, and six of 20 participants had a composite
score of exactly 5.00. This ceiling compression limited upward
discrimination and helps explain L5's departure from normality. Accordingly,
L5 can be described as receiving a high descriptive evaluation, but
not as demonstrably superior to L1 or L4 under the corrected comparisons.

As a secondary exploratory diagnostic, the coordinated multisensory
family (L4/L5: $M=4.24$, $SD=0.51$, median $=4.25$), static family (L1/L2:
$M=3.68$, $SD=0.65$, median $=3.83$), and L3 ($M=2.97$, $SD=0.67$,
median $=3.00$) differed overall, $\chi^2(2)=27.45$, $p<.001$, $W=.686$.
Bonferroni-adjusted comparisons found that the static family exceeded
L3 ($p_{\mathrm{adj}}=.002$, $r=.812$, HL = 0.75), the coordinated
multisensory family exceeded the static family ($p_{\mathrm{adj}}=.006$,
$r=.711$, HL = 0.58), and the coordinated multisensory family exceeded L3
($p_{\mathrm{adj}}<.001$, $r=.860$, HL = 1.33). Because these
families pool complete configurations that differ on several dimensions,
this diagnostic is not used as the paper's primary argument.

\subsection{Qualitative Results}
\label{sec:qual}

\subsubsection{Sensory Simplicity: ``Less but Clear'' as a Form of Comfort in Itself}

In the retrospective interviews, when the researcher asked
participants to compare their experience of L1 and L2 with the other
conditions, a recurring evaluative theme emerged: a moderate amount
of content and restrained sensory stimulation was itself actively
cited by a considerable number of participants as a reason for
liking the experience, rather than merely being a neutral state of
``no discomfort.''

\begin{quote}
``L1 was just plain text. Honestly, at first I thought it would be
boring, but after going through it, it actually felt the most
relaxing, because I knew nothing else was going to suddenly pop up,
and I could go through it at my own pace.'' [P1]
\end{quote}

\begin{quote}
``L2 just added pictures, nothing else jumped out to disturb me. I
really liked that `just right' feeling. It was quiet, and the space
and interface were understated, gentle, comfortable.'' [P6]
\end{quote}

\begin{quote}
``The fewer things there are, the easier it is for me to focus. L3
had richer content, but I actually remembered less from it than from
L2, because I kept having to process `what just happened.' My senses
got very tired that way, and I felt like I could not remember that
much.'' [P14]
\end{quote}

\begin{quote}
``Usually when I encounter information media in other museums, in
some noisy environments I have to work hard to filter out which
sounds matter and which do not. With L1 and L2 I did not have to do
that at all, the information itself was clean.'' [P8]
\end{quote}

For these participants, ``less'' was not described as an incomplete
experience. Sparse presentation was valued in its own right rather than
only serving as a baseline for comparison with multisensory conditions.

\subsubsection{Sensory Unity: The Integrated Comfort of Multiple Channels ``Telling the Same Story''}

In contrast to the ``subtraction'' logic of simplicity, participants
repeatedly mentioned another type of positive experience under
conditions L4 and L5. When the changes delivered through multiple
sensory channels were consistent with each other and jointly pointed
to the same narrative logic, the experience remained comfortable, or
even more enjoyable, even as the number of channels increased.

\begin{quote}
``In L5, when the light dimmed, the temperature on my arm dropped
along with it, and the sound was also describing the sperm whale
diving into the dark deep sea. These three things were telling the
same story, I did not need to separately figure out why the light
dimmed or why the temperature changed, they were supposed to change
together in the first place.'' [P14]
\end{quote}

\begin{quote}
``If I were just given a vibration on its own, unrelated to the
content, that would feel strange, like the device was misbehaving.
But in L5 the vibration followed the description of the heartbeat,
and instead I felt like this device `understood' what was being
described right then. Everything felt like it was happening
naturally, the way you would not be surprised by a bird flying
through the sky, or by a screen gradually turning dark in the deep
sea. All of it just felt normal.'' [P19]
\end{quote}

These accounts show that some participants did not experience L5's
additional channels as separate information points; they described the
cues as integrated in meaning and timing. This description is
consistent with L5 having the highest mean, but congruence was not
manipulated independently and cannot be identified as the cause of that
mean.

\subsubsection{Narrative Space, the Fragmented Feeling of the Interface: The Layering of Uncertainty and Loss of Control}

Participants' evaluations of L3 described a contrasting experience in
which uncertainty, suddenness, and reduced control were intertwined as
a ``fragmented experience.''

\textbf{The interweaving of uncertainty and suddenness.} Several
participants mentioned that the core problem with L3 was not the
intensity of the sound or image itself, but the lack of a sufficient
``warning signal'' before the trigger.

\begin{quote}
``Actually L3 does not have as much going on as L4 or L5, but the
moment I clicked, the sound just burst out. There was a small
speaker icon that gave a hint, but I still felt a jolt in my chest. I
feel like the sound was not necessary here, or rather, if there was
going to be sound, I should have been told in advance what would
happen.'' [P3]
\end{quote}

\textbf{The layered, fragmented experience.} Participants described their
sense of uncertainty as intensifying when several sensory events arrived
abruptly at the same time.

\begin{quote}
``I click once, and the image enlarges, a sound plays, and the
narration starts too, all three things come at once, I do not know
which one to pay attention to first. It is completely different from
the way L5 changes slowly and together. L3 is several things hitting
you at once, while L5 is one thing slowly unfolding. Actually I
already found the effect in L2 not that comfortable, the spatial
experience was not very clean, and L3 made it even more extreme.''
[P8]
\end{quote}

\textbf{The loss of a sense of control.} Although L3 used a
user-initiated trigger, several participants described control after
onset as incomplete.

\begin{quote}
``Once I click in L3, the sound just plays, and I cannot stop it
right away, I just have to wait for it to finish. In L4, when I turn
my head away, the effect slowly fades out, and it feels like the
initiative always stays with me, I can stop it at any time.'' [P13]
\end{quote}

Taken together, these accounts linked uncertainty, suddenness, and
reduced control within the ``fragmented feeling'' theme. Participants
described isolated events that could not be easily stopped after onset
and that positioned them reactively within the narrative.

\subsubsection{Individualized Spectrum of Sensory Sensitivity and Extended Situational Concerns}

Although the pattern above was fairly consistent at the group level,
the interviews also revealed clear individual differences.

P3's five composite scores were 4.33, 3.33, 3.00, 3.00, and 2.67 for
L1--L5, respectively, and correlated negatively with the five group means
(Spearman's $\rho=-.462$). This profile differed from the group-level
configuration ordering. At the same time, P3's earlier description of a
``jolt'' in L3 located one disliked moment at abrupt audio onset. Together,
the records show how one participant combined lower evaluations of several
configurations with a specific account of one disliked moment; they do not
separate channel quantity from onset dynamics as independent effects.

\begin{quote}
``The vibration part was actually fine for me, but I am especially
sensitive to temperature changes. During the later part of L5, when
the temperature dropped, my arm kept moving around, it felt a bit
uncomfortable, that was a point I did not really like.'' [P11]
\end{quote}

P11 nevertheless rated all three L5 items at the scale maximum
(5/5/5). This juxtaposition indicates that a global
post-condition score can mask a localized disliked moment. Momentary or
event-contingent ratings would be needed to capture such within-condition
variation directly.

Of particular note, a small number of participants showed
preferences opposite to the overall trend.

\begin{quote}
``Even though I gave the other conditions higher scores too, I did
not actually dislike L3, because I like knowing that `it only
responds because I clicked.' The automatic sensing in L4 and L5 made
me feel a bit unsafe, as if it would trigger on its own no matter
what I did, although this unease was not especially strong.'' [P17]
\end{quote}

In follow-up questioning, P17 distinguished distrust of automatic
system sensing from concern about what happened after clicking. This
case complicates any single account of uncertainty or control. It
generated a possible distinction between initiation locus (who starts
the event) and consequence predictability (what happens after onset),
but one divergent account cannot establish that these are independent
dimensions across participants.

\subsection{Integration of Quantitative and Qualitative Results}
\label{sec:integration}

Sections~\ref{sec:quant} and~\ref{sec:qual} address different parts of
the research questions and are integrated without collapsing their
evidential roles. First, the direct quantitative finding is that the five
complete configurations received different evaluations: L3 was rated lower
than the other configurations, while L4 and L5 received more positive
evaluations in several corrected comparisons. The narrower
Q1--Q2 analysis retained the omnibus difference, but corrected comparisons
did not distinguish L5 from L1 or L4 and L5 responses were
ceiling-compressed.
Second, the interview analysis records
participants' descriptions. Participants associated disliked moments in L3
with abrupt onset, simultaneous demands, insufficient warning, and limited
disengagement, and associated positive experiences in L4/L5 with gradual
transitions and coordinated narrative cues. These descriptions do not
explain the quantitative differences as tested mediators. Third, the
convergence between configuration-level results and participant accounts
motivates future controlled investigation of onset dynamics, cue
coordination, anticipation, and disengagement. Table~\ref{tab:joint-display}
keeps these three levels explicit.

\begin{table*}
  \caption{Joint display of direct quantitative findings, participant descriptions, and bounded interpretations}
  \label{tab:joint-display}
  \begin{tabular}{p{.28\linewidth}p{.28\linewidth}p{.36\linewidth}}
    \toprule
    Direct quantitative finding & Participant description & Bounded configuration-level interpretation \\
    \midrule
    L3 was tied for lowest for 16/20 participants; L4 exceeded L3 for 18/20; Q2 decreased by 0.85 from L2 to L3. &
    Participants associated disliked L3 moments with suddenness, simultaneous demands, insufficient warning, and reduced consequence control. &
    L3 received lower evaluations as a complete configuration. The participant accounts motivate controlled tests of onset, anticipation, and disengagement; they do not identify the cause of the configuration difference. \\
    L1--L2 Q2 differed by only 0.10, whereas Q1 and Q3 carried most of the composite decrease. &
    Some participants valued sparse and self-paced presentation for clarity and reduced filtering work. &
    Participants valued simplicity in these static configurations, but the comparison does not establish fewer channels as a general cause of comfort. \\
    L5 had the highest mean; L5 exceeded L1 for 14 participants versus one reversal and five ties; corrected L5--L1 and L5--L4 comparisons were not significant; 50\% of L5 item responses were at ceiling. &
    Participants associated positive L5 experiences with multiple cues that belonged to the same narrative event. &
    L5 received positive evaluations as a complete configuration. The accounts motivate controlled tests of cue coordination but do not establish coordination as the cause of those evaluations. \\
    P3's profile correlated negatively with the group means ($\rho=-.462$); P11 scored L5 at 5/5/5 while reporting a disliked thermal moment; P17 valued explicit initiation. &
    Participants described individualized sensitivities, localized disliked moments, and different preferences for initiation. &
    Case-level integration shows what global configuration ratings can hide and motivates moment-level measurement and personalization; it does not establish participant subgroups. \\
    \bottomrule
  \end{tabular}
\end{table*}

\section{Discussion}

\subsection{Five Prototypes as Probes of Preferred Sensory Expression}

We used five complete XR prototypes as concrete probes of preferred sensory
expression: what appeared, how cues unfolded and related to the narrative,
and how participants could act or disengage. The prototypes were not doses
of one variable. Direct quantitative findings show that L3 received lower
evaluations, while L4 and L5 received more positive evaluations in several
comparisons; they do not establish L5 as superior to L1 or L4. Interviews
then identify what participants valued or disliked:
P1 emphasized self-pacing, P14 and P19 related positive experiences to cues
serving one narrative, P3 and P8 described sudden or simultaneous demands,
and P13 valued being able to end an effect. The discussion therefore reports participant associations and uses them to qualify design assumptions and motivate controlled studies.

\subsection{What Participants Liked: Limits of Reduction-First Design}

P1 valued L1 because nothing would suddenly appear and the pace remained
personal; P6 called L2 quiet, gentle, and ``just right''; and P8 valued relief
from filtering sounds. These accounts support reducing unnecessary or
aversive stimulation~\cite{manning2023sensoryinclusive,
maclennan2023spiderweb}, while specifying that participants valued clarity,
self-pacing, and reduced filtering work.

Reduction was not a complete rule. P14 described L5's cues as ``telling the
same story,'' and P19 valued a vibration that followed the narrative rather
than appearing arbitrary. Thus, some participants did not experience every
additional, content-related cue as an additional burden. This qualifies a
design translation of sensory-reactivity research, not the evidence for
heterogeneous sensory differences itself~\cite{bensasson2009meta,
robertson2017sensory,tavassoli2018reactivity}. Sensory-inclusive guidance
also acknowledges an incomplete intervention evidence base
~\cite{manning2023sensoryinclusive} and interconnected effects of
sensoryscape, predictability, adjustment, and recovery
~\cite{maclennan2023spiderweb}.

For ISS, the contribution is participant-oriented evidence about complete
configurations: comfort was discussed through cue clarity, onset,
disengagement, and whether cues formed one narrative event. These accounts
define candidate dimensions, not a law relating sensory quantity to comfort.

\subsection{What Participants Disliked: Limits of Channel Count and
Initiation-as-Control}

L3 accounts challenge channel count as a proxy for burden. P3 noted that L3
had less going on than L4/L5 but disliked sound that ``burst out'' without
warning; P8 contrasted simultaneous demands in L3 with one event unfolding
in L5; and P14 described repeatedly determining ``what just happened.''
These statements do not identify the cause of L3's ratings, but show that
participants described timing, cue relations, and disengagement as well as
quantity. This aligns with work on sustained, inescapable, unpredictable,
or uncontrollable input~\cite{maclennan2023spiderweb}, feedforward, and
intolerance of uncertainty~\cite{djajadiningrat2002donald,
wensveen2004frogger,boulter2014uncertainty,neil2016uncertainty}.

Initiation also did not equal control. P13 could click to start L3 but not
stop it, whereas turning away ended L4. This fits relational accounts of
agency~\cite{coyle2012agency,limerick2014agency}. Conversely, P17 preferred
explicit initiation and distrusted automatic sensing, echoing gaze-based
Midas-touch concerns~\cite{jacob1990gaze,majaranta2002eyetyping}.
Initiation locus and post-onset control are therefore separate candidate
questions, not evidence that one trigger is universally preferable.

P11's ceiling-level L5 score coexisted with a disliked thermal moment;
P3 and P17 also diverged from the dominant accounts. These cases reinforce
documented sensory heterogeneity~\cite{robertson2017sensory,
tavassoli2018reactivity} and show why shared design priorities must still
allow individuals to adjust or leave an uncomfortable expression.

\subsection{Evidence-Graded Design Implications for Sensory-Considerate XR}

These candidates follow one chain: participant description, theoretical
convergence, then a dimension for controlled testing. Direction counts
compare complete prototypes and provide only indirect support.

\paragraph{Make the sensory expression legible.}
P1 and P8 valued self-pacing, clear expectations, and reduced filtering.
Feedforward and intolerance-of-uncertainty work likewise foreground
anticipation~\cite{djajadiningrat2002donald,wensveen2004frogger,
boulter2014uncertainty,neil2016uncertainty}. Legibility of content, action,
and system state is therefore a priority candidate. \emph{Evidence strength:
qualitative convergence only;} legibility was not independently manipulated
or isolated by a direction count.

\paragraph{Coordinate cues around a shared event.}
P14 and P19 valued cues belonging to one event, whereas P8 described
separate demands. Multisensory and crossmodal work similarly emphasizes
temporal and semantic relations~\cite{obrist2016touch,
spence2011crossmodal}. Cue coordination is therefore a priority candidate.
\emph{Evidence strength: qualitative convergence with indirect
prototype-level support;} L4 exceeded L3 for 18/20 participants and L5 was
tied highest for 17/20, but these counts do not isolate coordination.

\paragraph{Support anticipation, disengagement, and recovery.}
P3 wanted advance warning, and P13 valued being able to end an effect.
Feedforward, agency, and intolerance-of-uncertainty research converge on
anticipation and consequence control~\cite{djajadiningrat2002donald,
wensveen2004frogger,coyle2012agency,limerick2014agency,
boulter2014uncertainty,neil2016uncertainty}. Pre-onset signaling, staged
onset, and immediate pause or exit are therefore priority candidates.

\paragraph{Provide adjustment for individual sensory fit.}
P11, P17, and P3 showed localized or divergent responses. Agency and
sensory-inclusive research likewise emphasize adjustment and recovery
~\cite{coyle2012agency,limerick2014agency,maclennan2023spiderweb}.
Adjustable intensity, pacing, initiation, and exit are therefore priority
candidates. \emph{Evidence strength: qualitative and case-level evidence
only;} personalization was not manipulated, no direction count tested it,
and baseline sensory traits were not measured.

These implications prioritize future tests of legibility, coordination,
anticipation, disengagement, and adjustment; none is an established
independent effect.

\section{Limitations}

\subsection{Configuration and Session Scope}

Each L-level combined sensory channels, interaction, onset, pacing, and exit
affordances. Successive feature layering during construction was not a
controlled manipulation of sensory quantity; the findings concern five
complete configurations, and independent effects require component-controlled
studies. The 5 by 5 Latin square distributed order but could not remove
fatigue, novelty, or carryover within one session.

Repeating the same seven science segments controlled topic and wording but
may have introduced familiarity, habituation, boredom, anticipation,
contrast, or reduced processing demands. We did not measure repetition,
familiarity, boredom, recall, or condition-level fatigue, and prior content
or the preceding configuration may have influenced evaluations. Future work
should counterbalance matched content sets and measure these processes.

\subsection{Measurement and Generalizability}

The protocol-defined sensory comfort scale had pooled $\alpha=.895$,
but it combines two comfort-focused items with Q3's overall-liking
judgment and therefore is not a pure sensory-comfort measure. The Q1--Q2
sensitivity analysis retained the omnibus effect but changed some pairwise
conclusions. Condition-level reliability intervals were wide, and the
short, researcher-administered
scale reduced burden but may have introduced compliance or social-desirability
pressure and requires validation for autistic XR users. Friedman tests remain
primary; ANOVA is a labeled sensitivity analysis.

L5 had ceiling compression (30/60 maximum item responses; six of 20 composite
scores at 5.00), limiting discrimination and comparisons involving L5;
non-significant L1--L5 and L4--L5 tests do not establish equivalence. Baseline
sensory traits were not measured, so individual differences cannot be linked
to standardized sensory or clinical profiles.

The Chinese autistic adult sample required independent consent, diagnosis
documentation, social-media volunteering, and tolerance of visual, audio,
vibrotactile, and thermal XR. It likely favored lower-support-needs,
digitally literate participants and does not support population estimates
or generalization to all autistic people, autistic children, or people with
higher support needs~\cite{button2013power}. Without a
non-autistic group, the findings are not autism-specific. Translation from
Chinese adds interpretive risk despite bilingual checking.

\subsection{Qualitative and Exploratory Scope}

Analysts knew the quantitative results, which may have focused attention on
L3 and L5. The prototype-family summary is exploratory and descriptive, and
themes are researcher interpretations rather than participant-validated
categories because formal member checking was not conducted.

\section{Conclusion}

This study compared sensory comfort across five complete XR configurations
with autistic adults. Participants evaluated the configurations
differently: L3 received consistently lower ratings, while L4 and L5 were
evaluated more positively than L2 and L3 in corrected comparisons.
Corrected comparisons did not distinguish L5 from L1 or L4, and L5
ratings were ceiling-compressed. Participants described
valued experiences through clarity, self-pacing, gradual transitions,
disengagement, and coordinated narrative cues, while disliked experiences
were often described through abruptness, simultaneous demands, and limited
post-trigger control. These findings concern complete configurations rather
than the independent causal effects of individual sensory or interaction
components. Future controlled studies should isolate these
participant-described qualities and examine how repeated content exposure,
onset dynamics, interaction control, and multisensory organization shape
comfort.

\section{Data Availability, AI Use, and Accessibility}

The anonymized transcripts and diagnostic documentation cannot be
publicly released because the sample is small and the interviews
include potentially identifiable sensory and health information.
The de-identified item-level comfort scores and reproducible analysis
script can be shared upon reasonable request, subject to the study's
ethics and data-governance requirements.

Generative AI tools were used for language editing and reference
formatting support during manuscript preparation. The authors remain
responsible for the accuracy of claims, analysis, citations, and final
text.

Figures include text descriptions through the ACM \texttt{Description}
field.
\bibliographystyle{ACM-Reference-Format}
\bibliography{reference}

\appendix
\section{Final Thematic Framework}
\label{app:codebook}

This appendix records the final thematic framework produced through the
reflexive thematic analysis. The themes and boundaries were refined through
repeated engagement with the transcripts and negotiated discussion between
the two researchers; they were not predetermined coding categories.

\subsection{Theme 1: Sensory Simplicity}

\textbf{Definition.} Experiences in which participants valued reduced
sensory complexity, clear information presentation, and control over their
own pace.

\textbf{Indicative initial codes.} Visual clarity; self-paced reading;
reduced sensory demands; ease of concentration; preference for static
content.

\textbf{Analytic boundary.} Included statements described the comfort,
clarity, or manageability of static text or text-and-image presentations.
Statements primarily about unexpected onset, inability to interrupt a cue,
or coordination among multiple channels were excluded. The theme was most
closely associated with L1 and L2.

\subsection{Theme 2: Fragmented Experience of L3}

\textbf{Definition.} Experiences in which abrupt sensory events,
insufficient warning, and limited opportunities to disengage disrupted
participants' attention or sense of control.

\textbf{Indicative initial codes.} Suddenness of trigger; lack of warning
signals; interrupted attention; limited interruptibility; loss of control;
difficulty anticipating sensory events.

\textbf{Analytic boundary.} Included statements described abrupt audio
onset, discomfort after clicking, interrupted attention, or difficulty
stopping a triggered event. General dislike of audio without reference to
onset, timing, anticipation, or disengagement was excluded. The theme was
primarily associated with L3.

\subsection{Theme 3: Sensory Unity}

\textbf{Definition.} Experiences in which multiple sensory cues were
perceived as coherent because they followed a shared narrative event and
were introduced in a gradual, coordinated manner.

\textbf{Indicative initial codes.} Coordinated sensory cues; narrative
coherence; gradual transitions; predictable sensory onset; integration of
sensory information; opportunities to disengage.

\textbf{Analytic boundary.} Included statements described gradual sensory
transitions, coordinated changes across modalities, or relationships between
cues and narrative content. General enjoyment of multisensory stimulation
without describing its organization or coherence was excluded. The theme
was primarily associated with L4 and L5.

\subsection{Theme 4: Individualized Spectrum of Sensory Sensitivity}

\textbf{Definition.} Individual differences in sensory preferences,
tolerance, and responses to the same XR configurations, including
experiences that diverged from the group-level pattern.

\textbf{Indicative initial codes.} Individual sensory preferences;
differences in sensory tolerance; divergent comfort evaluations; personal
sensitivity to specific modalities; preference for particular interaction
styles.

\textbf{Analytic boundary.} Included statements indicated that sensory
comfort depended on individual needs, sensitivities, or preferences,
including accounts that differed from the dominant group pattern. General
descriptions of a configuration without evidence of individual variation
were excluded. The theme applied across all five configurations.

\subsection{Resolution of Interpretive Differences}

The researchers compared their initial codes and repeatedly examined each
participant's surrounding account when interpretations differed. Codes for
sudden onset, insufficient warning, and limited control, for example, were
initially separate and were later understood as interconnected aspects of
the fragmented experience associated with L3. Theme boundaries were refined
through negotiated consensus rather than an intercoder reliability
coefficient, consistent with the interpretive orientation of reflexive
thematic analysis.

\section{Hardware and Stimulus Settings}
\label{app:hardware}

Table~\ref{tab:hardware-settings} summarizes the implementation settings.
The approximately two-second values refer to activation dwell and the
visual-auditory transition. Thermal cues followed their own controlled ramp
rate and were coordinated with the relevant narrative event rather than
completing the full temperature change within two seconds.

\begin{table*}[t]
  \caption{Hardware and reported stimulus settings}
  \label{tab:hardware-settings}
  \begin{tabular}{p{.29\linewidth}p{.65\linewidth}}
    \toprule
    Component or parameter & Setting \\
    \midrule
    XR headset & Meta Quest 3 standalone XR headset \\
    Display refresh rate & 90 Hz \\
    Spatial audio & Headset-integrated stereo spatial audio \\
    Audio level & Approximately 65--70 dBA \\
    Vibrotactile device & Forearm-mounted vibration actuator \\
    Vibration setting & Approximately 150 Hz at 20\% of maximum actuator output \\
    Thermal device & Forearm-mounted heating and cooling patch \\
    Thermal safety envelope & 28--36$^\circ$C \\
    Thermal stimulus range & 30--34$^\circ$C \\
    Temperature change rate & Approximately 0.3$^\circ$C/s \\
    L3 audio onset & Immediate onset after click, without programmed fade-in \\
    L4/L5 activation & Approximately two-second gaze/proximity dwell \\
    L4/L5 visual-auditory transition & Approximately two-second gradual fade-in \\
    L5 vibrotactile and thermal timing & Initiated in coordination with the corresponding narrative event, using channel-specific dynamics \\
    Exit mechanism & Turning or moving away initiated fade-out in L4/L5; participants could request a session-level interruption at any time \\
    Safety controls & Temperature monitoring, configured output limits, automatic thermal shutdown, and researcher- or participant-initiated session interruption \\
    \bottomrule
  \end{tabular}
\end{table*}

\end{document}